# Finite Difference Time Domain Method for Computing Single-Particle Density Matrix

I Wayan Sudiarta [1,*] and D J Wallace Geldart [1,2]

[1] Department of Physics and Atmospheric Science, Dalhousie University, Halifax, NS B3H 3J5, Canada

[2] School of Physics, University of New South Wales, Sydney, NSW 2052, Australia

E-mail : sudiarta@dal.ca

**Abstract**

A general method for numerical computation of the thermal density matrix of a single-particle quantum system is presented. The Schrödinger equation in imaginary time $\tau$ is solved numerically by the finite difference time domain (FDTD) method using a set of initial wave functions at $\tau = 0$. By choosing this initial set appropriately, the set of wave functions generated by the FDTD method can be used to construct the thermal density matrix. The theoretical basis of the method, a numerical algorithm for its implementation, and illustrative examples in one, two and three dimensions are given in this paper. The numerical results show that the procedure is efficient and accurately determines the density matrix and thermodynamic properties of single-particle systems. Extensions of the method to more general cases are briefly indicated.

## 1. Introduction

A quantum system can be fully described by a single state vector only if the system is in a pure state. Isolated atoms or molecules in eigenstates of their Hamiltonian are among the common examples of such systems. In practice, a great many physical systems of interest are not in pure states but instead are in mixed states. This requires a description in terms of a statistical operator or density matrix [1]. Time-independent systems in thermal equilibrium with a heat bath at constant temperature $T$ are important practical examples of mixed states requiring a density matrix description.

The density matrix of a system in thermal equilibrium can be specified by its matrix elements in the position representation [1],

$$\rho(R, R', \beta) = \sum_{n=0}^{\infty} \phi_n(R) \phi_n^*(R') \rho_n \qquad (1)$$

---

* Corresponding Author. E-mail: sudiarta@dal.ca

Coordinate vectors are denoted by $R$ and spin degrees of freedom will be suppressed for simplicity. $E_n$ and $\phi_n(R)$ are the energy eigenvalues and eigenfunctions of the time-independent Schrödinger equation $\hat{H}\phi_n(R) = E_n\phi_n(R)$, $\beta = 1/k_B T$ with $k_B$ the Boltzmann constant, $\rho_n = [\exp\{-\beta E_n\}]/Z(\beta)$ and $Z(\beta) = \sum_{n=0}^{\infty} \exp\{-\beta E_n\}$ is the partition function.

Solving the Schrödinger equation to obtain all energy eigenvalues and wave functions of a general many-particle system is obviously not feasible. Analytical solutions of the Schrödinger equation are available only for a relatively small number of idealized cases. For practical applications, alternative numerical methods are needed to determine the density matrix, even for one-particle systems. The most effective numerical procedure for a particular problem depends on the number of particles $N$ in the system and the spatial dimension $D$. We will comment briefly on procedures currently available and then indicate our proposed finite difference method.

For systems of large particle number, the path integral Monte Carlo (PIMC) method [2, 3] is very effective due to the high efficiency of stochastic methods for sampling high dimensional configuration spaces when computing multidimensional integrals [4]. However, there are two disadvantages of the PIMC method. The treatment of nodes of many-fermion wave functions is approximate (the sign problem) [5]. Also, convergence the PIMC method is rather slow since variances are of order $1/\sqrt{N_C}$ where $N_C$ is the number of sampled configurations. Accurate results then require averaging over a large number of configurations with correspondingly long simulation times. In spite of these disadvantages, PIMC is still indispensable for large particle number, particularly for higher spatial dimension, due to the efficiency of stochastic methods.

In the opposite extreme of small particle number, alternative computational procedures become practical. These include the density matrix renormalization group (DMRG) method [6-8], the discretized path integral (DPI) method [9,10], the finite difference time domain (FDTD) method [11-13], the Lanczos method [14,15] and the Chebyshev expansion method [16]. These have the potential for improved accuracy over the PIMC method since they are not based on slowly convergent stochastic sampling.

For a system of particles on a strictly one dimensional lattice, the DMRG method gives much better accuracy than Monte Carlo methods, even when the number of particles is not small. The method can be applied to higher $D$ by mapping to a one-dimensional model with long range interactions. Although such models can be computationally feasible, the memory requirements increase rapidly due to the need to keep a much larger number of basis states. General applications of the DMRG method to two- and three-dimensional systems are still in development [7].

The DPI method for computing the density matrix requires manipulation of a density matrix with $N_g \times N_g$ elements (where $N_g$ is the number of grid points). The density matrix can also be computed by converting Eq. (1) to a differential equation (Bloch equation) [17,18]. This well known Bloch equation can be solved by a variety of methods, all of which require manipulation of a matrix with $N_g \times N_g$ elements, just as in the DPI method.

The Lanczos method uses a set of random initial wave functions. For each wave function, the Lanczos procedure [19] is performed for a finite number of iterations resulting in a finite set of eigenvalues and eigenfunctions. These eigenvalues and eigenfunctions are then used to approximate the density matrix. The drawbacks of this method are the loss of accuracy due to the use of random wave functions and the loss of orthogonality when the number of Lanczos iterations is increased.

As an alternative method for integrating the Schrödinger equation, one can use the Chebyshev expansion method or the FDTD method. Depending on the number of terms in the Chebyshev expansion, one can solve the Schrodinger equation with larger time step than in the FDTD method. However for each time step, the number of numerical operations is larger than for the FDTD method.

The purpose of this paper is to present an efficient and robust general purpose procedure for the numerical computation of the density matrix for inhomogeneous few-particle systems. We first show that the density matrix can be constructed from a set of wave functions obtained by integrating the Schrodinger equation in imaginary time, starting from a suitably chosen set of initial functions at $\tau = 0$. This use of individual wave functions, rather than operating with the matrix itself, results in an efficient alternative to the DPI method since computation with a discretized wave function on a grid requires only $N_g$ elements as compared to $N_g \times N_g$ elements in the DPI or other matrix–based methods.

It is important to emphasize that the chosen initial wave functions are not the energy eigenfunctions of the system. In fact, no information on the exact $\{E_n\}$ and $\{\phi_n(R)\}$ is required. The procedure for selecting the set of initial wave functions to be used and the proof that the resulting construction is equivalent to the conventional definition of the density matrix Eq. (1) are given in section 2. This proof is valid for any $N$ and $D$, and for general one-body potentials and interparticle interactions.

We are mainly interested in quasi-two-dimensional few-electron systems, such as low density heterojunctions and transistor devices or quantum dots. In addition to the confining potential, the system will generally have quenched disorder due to a fixed configuration of defects, such as impurities. The overall inhomogeneity is due to the one-body confining potential or the one-body potential of the quenched impurities, or to both. The major common computational feature shared by these systems is the need to treat reasonably general spatial variations of the total one-body potential on nanometer or possibly atomic scales. There are no symmetries or other simplifying features. In the case of random disorder, it may also be necessary to average over many configurations of the impurities.

The numerical method chosen for integrating the Schrödinger equation must be robust and adaptable, easily applicable to one-body potentials of the expected physical type, and of an accuracy (to be determined by the specific problem at hand) which can be well controlled by the user. The finite difference time domain (FDTD) method has previously been used to solve time-independent and time-dependent Schrodinger equations to obtain energy eigenfunctions and eigenvalues for a variety of one-body potentials [11-13,20]. The scale of the discrete finite difference grid can always be chosen to reflect the spatial variation of the potential in any specific problem at hand so the inhomogeneity can be accomodated. The FDTD method has been applied also to the scattering of electromagnetic waves by heterogeneous particles [21]. This is an electromagnetic analogue of the disordered electron systems which we wish to consider. In all of these cases, the FDTD method has been found to be flexible, robust and accurate for the applications considered.

The extension of the FDTD method to direct computation of the thermal density matrix for a general single-particle system, without use of energy eigenfunctions and eigenvalues, is given in section 3. An easily implementable numerical algorithm is also given. We do not consider two-body interactions at this point. In fact, the problem of dealing with generic inhomogeneity is present *irrespective* of such interactions. Consequently, the usefulness of the FDTD procedure for computing the density matrix for inhomogeneous systems can be sufficiently verified by a study of representative single-particle examples. Accordingly, subsequent discussion and numerical examples are given for the single-particle $N = 1$ case in the remainder of this paper. However, as will be seen from section 2, the analytical framework applies also to systems of interacting particles. We have verified that the FDTD numerical procedures can be extended to $N \geq 2$ cases, including two-body interactions. Of course, additional features do appear. A brief discussion of this is given in the conclusions and details will be given elsewhere.

Numerical results for a variety of single-particle model systems are given in section 4. We first consider one-dimensional examples so that the accuracy of density matrices and thermodynamic properties computed by the FDTD method can be easily judged by comparison to either analytical results or other numerical procedures. The infinite square well and the harmonic oscillator are two standard test cases where analytical results are known. Convergence properties of our method are good and our numerical results are in excellent agreement with the analytical results. Next, density matrices and thermodynamic properties are computed for a wide range of temperature for a linear confining potential, a quartic oscillator, and a double well potential in one dimension. Convergence properties and accuracy are again good. Results have not previously been known for these cases. We then consider a two-dimensional model of a double quantum dot, including the effect of an applied electric field. Finally, a quartic oscillator is considered in three dimensions. The primary purpose of these tests is the validation of the general numerical procedure for a variety of model potentials to simulate spatial inhomogeneity. However, the physical relevance of examples is indicated when appropriate.

Finally, in section 5 we summarize our conclusions. We also comment on extensions of the present FDTD method to more general systems of two or more particles.

**2. Theory**

In this section we give a general procedure for constructing the matrix elements $\rho(R, R', \beta)$ for a quantum system in thermal equilibrium. We begin by considering the Bloch density matrix [18]

$$C(R, R', \beta) = \sum_{n=0}^{\infty} \phi_n(R) \phi_n^*(R') \exp\{-\beta E_n\} \qquad (2)$$

We will obtain $C(R, R'; \beta)$ *without* the need to have all the energy eigenvalues and eigenfunctions, $E_n$ and $\phi_n(R)$. The procedure has two steps. We solve the time-dependent Schrödinger equation in imaginary time with selected initial conditions. The Bloch density matrix is then formed by summing products of these wave functions for an appropriate set of initial conditions.

A general solution of the time-dependent Schrödinger equation

$$i\hbar \frac{\partial}{\partial t}\psi(R,t) = \hat{H}\psi(R,t) \qquad (3)$$

can be expanded in the set of energy eigenfunctions $\{\phi_n(R)\}$

$$\psi(R,t) = \sum_{n=0}^{\infty} c_n \phi_n(R) \exp(-iE_n t/\hbar) \qquad (4)$$

where $\{c_n\}$ is a set of expansion coefficients.

In imaginary time $\tau = it/\hbar$, Eqs. (3) and (4) become

$$\frac{\partial}{\partial \tau}\psi(R,\tau) = -\hat{H}\psi(R,\tau) \qquad (5)$$

and

$$\psi(R,\tau) = \sum_{n=0}^{\infty} c_n \phi_n(R) \exp(-E_n \tau) . \qquad (6)$$

The solution of Eq. (5) generates $\psi(R,\tau)$ from the specified initial wave function $\psi(R,\tau=0)$. Therefore, we can generate a set of $\{\psi_k(R,\tau)\}$ corresponding to a set of independent initial wave functions $\{\psi_k(R,\tau=0)\}$. Consequently, we can also form a set of products $\{\psi_k(R,\tau)\psi_k(R',\tau)^*\}$. Note that the required $\psi_k(R,\tau)^*$ is simply the complex conjugate of $\psi_k(R,\tau)$ which is computed using Eq. (5). According to Eq. (6) the required product is

$$\psi_k(R,\tau)\psi_k(R',\tau)^* = \sum_{n=0}^{\infty}\sum_{m=0}^{\infty} c_{k,n} c_{k,m}^* \phi_n(R)\phi_m^*(R') \exp\{-\tau(E_n + E_m)\} \qquad (7)$$

The crucial step to form the density matrix is to choose a complete set of orthonormal wave functions $\{\chi_k(R)\}$ for the initial wave functions $\psi_k(R,\tau=0)$.

The density matrix will then be obtained by summing the set of products $\{\psi_k(R,\tau)\psi_k(R',\tau)^*\}$,

$$\sum_{k=0}^{\infty} \psi_k(R,\tau)\psi_k(R',\tau)^* = \sum_{k=0}^{\infty}\sum_{n=0}^{\infty}\sum_{m=0}^{\infty} c_{k,n} c_{k,m}^* \phi_n(R)\phi_m^*(R') \exp\{-\tau(E_n + E_m)\} \qquad (8)$$

To see that Eq. (7) is indeed the Bloch density matrix, note that its value at $\tau=0$ is

$$\sum_{k=0}^{\infty} \psi_k(R,\tau=0)\psi_k(R',\tau=0)^* = \sum_{n=0}^{\infty}\sum_{m=0}^{\infty}\left[\sum_{k=0}^{\infty} c_{k,n} c_{k,m}^*\right] \phi_n(R)\phi_m^*(R') \qquad (9)$$

But the left hand side is

$$\sum_{k=0}^{\infty} \chi_k(R)\chi_k^*(R') = \delta(R-R') \qquad (10)$$

Then the right hand side of Eq. (9) must also be $\delta(R-R')$. This requires

$$\sum_{k=0}^{\infty} c_{k,n} c_{k,m}^* = \delta_{n,m} \qquad (11)$$

Using this in Eq. (8) yields

$$\sum_{k=0}^{\infty} \psi_k(R,\tau)\psi_k(R',\tau)^* = \sum_{n=0}^{\infty} \phi_n(R)\phi_n^*(R')\exp(-2\tau E_n) \qquad (12)$$

which demonstrates that

$$C(R,R';\beta) = \sum_{k=0}^{\infty} \psi_k(R,\tau=\beta/2)\psi_k(R',\tau=\beta/2)^* \qquad (13)$$

The partition function is obtained by

$$\int C(R,R,\beta)dR = \sum_{n=0}^{\infty} \exp\{-\beta E_n\} = Z(\beta) \qquad (14)$$

using $\int \phi_n(R)\phi_n^*(R)dR = 1$. The canonical thermal density matrix is then given as $\rho(R,R';\beta) = C(R,R';\beta)/Z(\beta)$. We emphasize that this exact construction of the density matrix requires no knowledge of the stationary state energy eigenvalues or eigenfunctions of $\hat{H}$.

There are two issues involved in constructing the density matrix by means of Eq. (13). One is the choice of procedure for numerical integration of Eq. (5) and the second is the choice of the set of initial functions $\{\chi_k(R)\}$ to be used in Eq. (13). These issues are discussed in the following section.

## 3. Numerical Procedure

It was shown in the previous section that the density matrix can be constructed by computing an appropriate set of wavefunctions at imaginary time $\tau$. For any chosen initial wave function $\psi(R,\tau=0)$, one can solve Eq. (5) numerically to obtain the evolved wave function at $\tau > 0$. As discussed in the introduction we use the finite difference time domain (FDTD) method given in our previous paper [12]. This method was used to solve Schrodinger equations in imaginary time to obtain energy eigenvalues and eigenfunctions for a variety of systems in confining potentials in one, two, and three dimensions. The cases studied in [12] were of same type as considered here for density matrix applications and so the same requirements for performance and stability apply. We found the FDTD method to be flexible and easily implemented, robust, and of good accuracy for these inhomogeneous systems. Of course, other integration methods could be considered for specific cases. For example, integration routines based on spectral or pseudo-spectral methods are useful in situations where memory constraints are acute. However when the confining potentials are rapidly varying or irregular, as will be the case for applications when quenched impurities are present, the global spectral methods tend to be less efficient than finite difference algorithms [22]. Consequently the well tested FDTD method will be used in this work.

As discussed in [12], a finite computational volume is adopted in our FDTD method. The space is truncated at an outermost boundary by imposing the boundary condition $\psi(x,y,z,\tau)_{boundary} = 0$. This boundary condition at a finite distance does not affect the results significantly provided that the simulation space is chosen to be large enough so that wave functions have already decayed to sufficiently small values at the boundary. The size of the computational volume is to be chosen accordingly in each application. Other boundary condition such as periodic boundary and open boundary conditions may also be used in some cases. In our applications to an isolated system or subsystem, the zero boundary condition is sufficient.

The computational domain will be discretized. For example, in three dimensions we take $(N_x +1) \times (N_y +1) \times (N_z +1)$ grid points with the grid spacings given by $\Delta x = L_x / N_x, \Delta y = L_y / N_y$ and $\Delta z = L_z / N_z$. The grid positions are $(i\Delta x, j\Delta y, k\Delta z)$. A notation $\psi^n(i,j,k) = \psi(i\Delta x, j\Delta y, k\Delta z, n\Delta\tau)$ is used where $\Delta\tau$, $\Delta x$, $\Delta y$ and $\Delta z$ are the temporal and spatial grid spacings with i, j and k integers. Since $\psi(x,y,z) = 0$ at the outer boundary, there are $(N_x -1) \times (N_y -1) \times (N_z -1)$ undefined variables at the remaining internal grid points. The initial wave functions $\{\chi_k(R)\} = \psi_k(R,\tau=0)\}$ in the interior of the computational space can be represented by these $(N_x -1) \times (N_y -1) \times (N_z -1)$ variables. This is the maximum number of initial functions that can be represented on this grid. With this discretization procedure Eq. (5) can then be integrated numerically for a specific application by the algorithm given in our previous paper [12].

$$\psi^{n+1}(i,j,k) = a\psi^n(i,j,k) + b \left[ \begin{array}{l} \frac{\Delta\tau}{2\Delta x^2}\left[\psi^n(i+1,j,k) - 2\psi^n(i,j,k) + \psi^n(i-1,j,k)\right] \\ + \frac{\Delta\tau}{2\Delta y^2}\left[\psi^n(i,j+1,k) - 2\psi^n(i,j,k) + \psi^n(i,j-1,k)\right] \\ + \frac{\Delta\tau}{2\Delta z^2}\left[\psi^n(i,j,k+1) - 2\psi^n(i,j,k) + \psi^n(i,j,k-1)\right] \end{array} \right]$$

(15)

where the coefficients $a$ and $b$ are given by

$$a = \frac{[1-\frac{\Delta\tau}{2}V(i,j,k)]}{[1+\frac{\Delta\tau}{2}V(i,j,k)]}, \quad b = \frac{1}{[1+\frac{\Delta\tau}{2}V(i,j,k)]} \tag{16}$$

For a stable computation the time step $\Delta\tau$ must satisfy a stability condition given by
$\Delta\tau \leq [\frac{1}{\Delta x^2} + \frac{1}{\Delta y^2} + \frac{1}{\Delta z^2}]^{-1}$.

Equation (15) is used iteratively to evolve the wave function for each member of the initial set. The Bloch density matrix is then obtained from Eq. (13).

We now examine the issue of constructing a set of computationally simple and efficient initial functions which are orthonormal on the discrete grid. Consider first a set of $(N_x -1) \times (N_y -1) \times (N_z -1)$ localized initial functions defined by $\chi_k(R) = 1$ if $R$ is inside the cell specified by $(i,j,k)$ and $\chi_k(R) = 0$ otherwise. This set is clearly orthonormal on the discrete grid. Sets of extended initial functions can then be formed by taking linear combinations of these localized functions. The results reported in this paper were obtained with linear combinations corresponding to digitization of sine functions,

$$\chi_{uvw}(x,y,z) = \sqrt{\frac{8}{L_x L_y L_z}} \sin(\frac{u\pi x}{L_x}) \sin(\frac{v\pi y}{L_y}) \sin(\frac{w\pi z}{L_z}) \tag{17}$$

where $u$, $v$ and $w$ are integers. These functions satisfy $\psi(x,y,z) = 0$ at the outer boundary and are orthonormal in the computational volume. Also it is convenient that the ordering and spacing in energy of the initial functions is known exactly. Then since the number of sine functions that can be represented in the computational cell is $(N_x -1) \times (N_y -1) \times (N_z -1)$, the maximum energy in the discrete computational domain is approximately

$$E_{max} = \frac{\hbar^2 \pi^2}{2m} \left[ \left(\frac{N_x -1}{L_x}\right)^2 + \left(\frac{N_y -1}{L_y}\right)^2 + \left(\frac{N_z -1}{L_z}\right)^2 \right] \tag{18}$$

This corresponds to a maximum temperature $T_{max} = E_{max}/k_B$. Numerical results for the density matrix will be accurate only if the temperature of the physical system $T$ is significantly smaller than this $T_{max}$.

It is clear that the accuracy of the computed density matrix depends on the number of initial functions used. The level of accuracy required for a specific physical problem can always be achieved by increasing the number of grid points and the number of initial wave functions on the grid. It is important that the numerical procedure we propose should converge rapidly as the number of initial functions is increased. Then a relatively small number of initial functions will give an accurate density matrix and thermodynamic properties for the temperature range of interest. Examples of this are given in the following section.

Finally we note that there are alternative procedures for choosing the set of initial functions. One could use a set of random wave functions as shown by Hams and De Raedt [23]. However the numerical errors due to random initial functions are of order $1/\sqrt{M}$ (where $M$ is the number of initial wave functions) so convergence may be slow. The use of orthogonal initial functions is more efficient.

Computer memory requirement of the FDTD method is dependent on the number of grid points in the computational domain. For example, for a three dimensional system, the amount of memory is proportional to $N_x \times N_y \times N_z$. If the DPI method is used the amount of memory is quadratic in $N_x \times N_y \times N_z$. The CPU time for computing the density matrix is proportional to the product of the number of iterations $N_\tau$ and the number of initial wave functions $N_\psi$. The number of iterations is proportional to $\beta = T^{-1} = 2N_\tau \Delta\tau$. Then for low temperature system a large number of iterations is required. However, it will be shown in the next section that for low temperature systems, only a small number of initial wave functions is needed which results in significant computational efficiency. Moreover, the computation speed can be improved further by parallel compution since the iterations for every initial wave function can be performed independently.

**4. Numerical Results and Discussion**

In this section we give numerical results for the density matrix and for the thermodynamic properties for representative single-particle model systems. We first give results for the one-dimensional infinite square well and the one-dimensional harmonic oscillator. Exact analytical results are known for these two standard test cases. Our numerical results are in excellent agreement with the exact results. This illustrates the general procedures for accuracy control and the convergence properties of the method. We then consider examples where exact analytical results, and often not numerical results, for the density matrix and thermodynamic properties have not previously been known. The density matrix and thermodynamic properties as a function of temperature are then computed for a linear confining potential, a quartic oscillator and a double-well potential, still all one-dimensional. Next we treat a two-dimensional double quantum dot, including an applied electric field. Finally we consider a quartic oscillator in three dimensions. These examples give good tests of our FDTD procedure for a variety of model inhomogeneous single-particle systems. Comments on other cases are given in the conclusions.

To express computations in dimensionless form, we use units where $\hbar$, $m$, $k_B$ and a length scale specified by the potential are all set equal to unity. Then temperature is measured in energy units and the entropy will be dimensionless.

**4.1. One-dimensional square well**

As a first example, we consider a particle in a one-dimensional infinite square well with a dimensionless width $a = \pi$; $V(x) = 0$ for $0 < x < \pi$ and $V(x) = \infty$ for $x > \pi$ and $x < 0$. The energy eigenfunctions for this problem are

$$\psi_n(x) = \begin{cases} \sqrt{2/\pi} \sin(nx), & \text{for } 0 < x < \pi \\ 0, & \text{for } x < 0 \text{ or } x > \pi \end{cases} \tag{19}$$

and the eigenvalues are $E_n = n^2/2$. The Bloch density matrix can be constructed in the basis of energy eigenfunctions as

$$C(x, x', \beta) = \begin{cases} \dfrac{2}{\pi} \sum_{n=1}^{\infty} \exp(-n^2 \beta/2) \sin(nx) \sin(nx'), & \text{for } 0 < x < \pi \text{ and } 0 < x' < \pi \\ 0, & \text{otherwise} \end{cases} \quad (20)$$

In principle, the right hand side of Eq. (20) can be evaluated by direct summation to any required accuracy. However the convergence is very slow at high temperature. To improve convergence this representation can conveniently be expressed exactly in terms of Jacobi theta functions [24] as

$$C(x, x', \beta) = \begin{cases} \dfrac{1}{2\pi} [\Theta_3((x - x')/2, q) - \Theta_3((x + x')/2, q)] & \text{for } 0 < x < \pi \text{ and } 0 < x' < \pi \\ 0, & \text{otherwise} \end{cases} \quad (21)$$

where $q = \exp(-\beta/2)$. Application of the Jacobi imaginary transformation then yields an alternative series representation for the Bloch density matrix that converges rapidly at high temperature. Very accurate results are then available at arbitrary temperature for comparison with our numerical method.

Numerical results are shown in Fig. 1 to 3. In this application, the parameters $\Delta x = 0.02\pi$, $\Delta \tau = 0.001$ are used. The numerical results in Figs. 1 to 3 are in an excellent agreement with the known exact results. The contour lines in Fig. 1 for the computed density matrix $\rho(x, x', \beta)$ coincide with the exact contour lines obtained from Eq. (21) on this scale. This good agreement is also clear in the plot given in Fig. 2 for the particle density, the diagonal elements of the density matrix $n(x) = \rho(x, x, \beta)$. Furthermore, numerical results for the free energy in Fig. 3 are seen to agree well with the exact results over the wide temperature range considered (0.5 to 45).

Figs. 2 and 3 also show the good convergence of the numerical results to the exact results as the number of initial wave functions increases. It is very important that accurate results can be obtained by using only a small number of initial wave functions. For example, the numerical results in Fig. 2b coincide (within less than 0.1%) with the exact results even though only four initial wave functions are included. This illustrates that accurate results for a specified range of $T$ of interest are guaranteed by a relatively small number of initial orthonormal functions provided that the corresponding energy range is well represented by these functions.

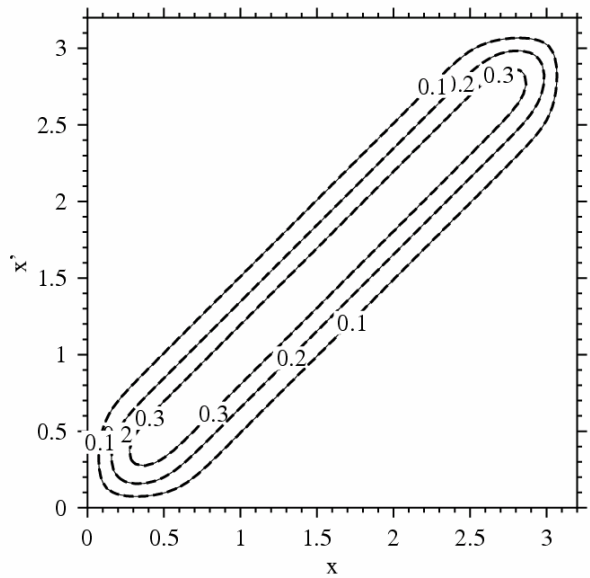
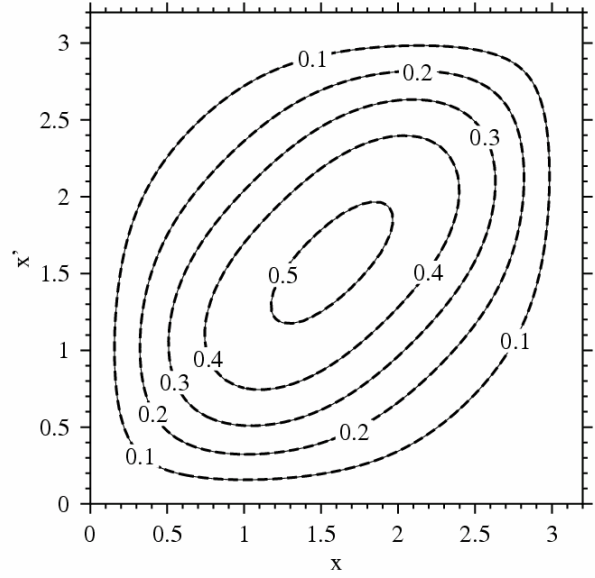

(a)            (b)

**Figure 1.** A contour plot of numerical results (heavy dashed lines) for the density matrix $\rho(x, x', \beta)$ compared with the exact solution (solid lines) for a particle in an infinite square well. The contour lines for numerical results essentially coincide with the exact solution on this scale. The parameter $\beta$ is $\beta = 0.1$ (or $T = 10$) for (a) and $\beta = 1$ (or $T = 1$) for (b). Good agreement is also obtained for other values of T. The number of initial functions used is 49 for (a) and only 4 for (b).

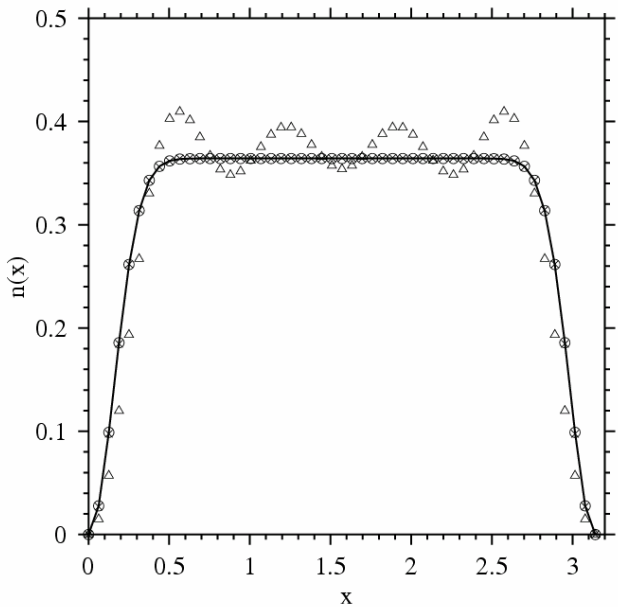
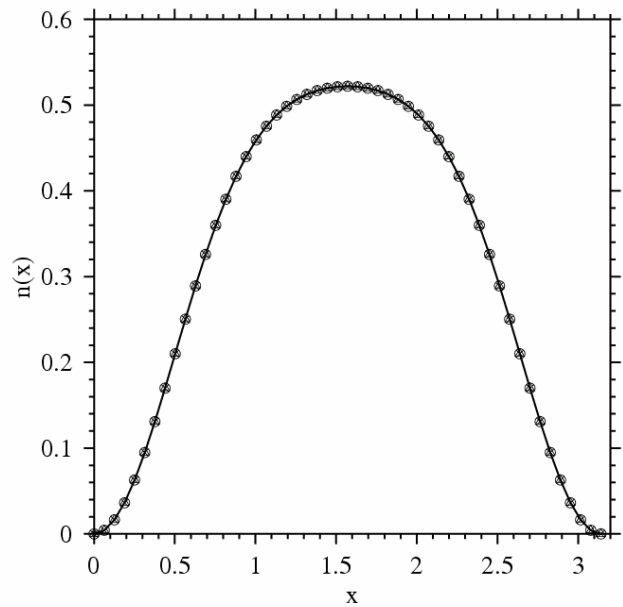

(a)            (b)

**Figure 2.** A comparison of numerical results for the diagonal elements of the density matrix $n(x) = \rho(x,x,\beta)$ with the exact solution (solid line) for a particle in an infinite square well at $\beta = 0.1$ (or $T = 10$) for (a) and $\beta = 1$ (or $T = 1$) for (b) as in Fig. 1. The circles, crosses, and triangles correspond to 49, 9 and 4 initial wave functions respectively. The results indicate good convergence with respect to the number of low-energy sine functions in the initial wave function set.

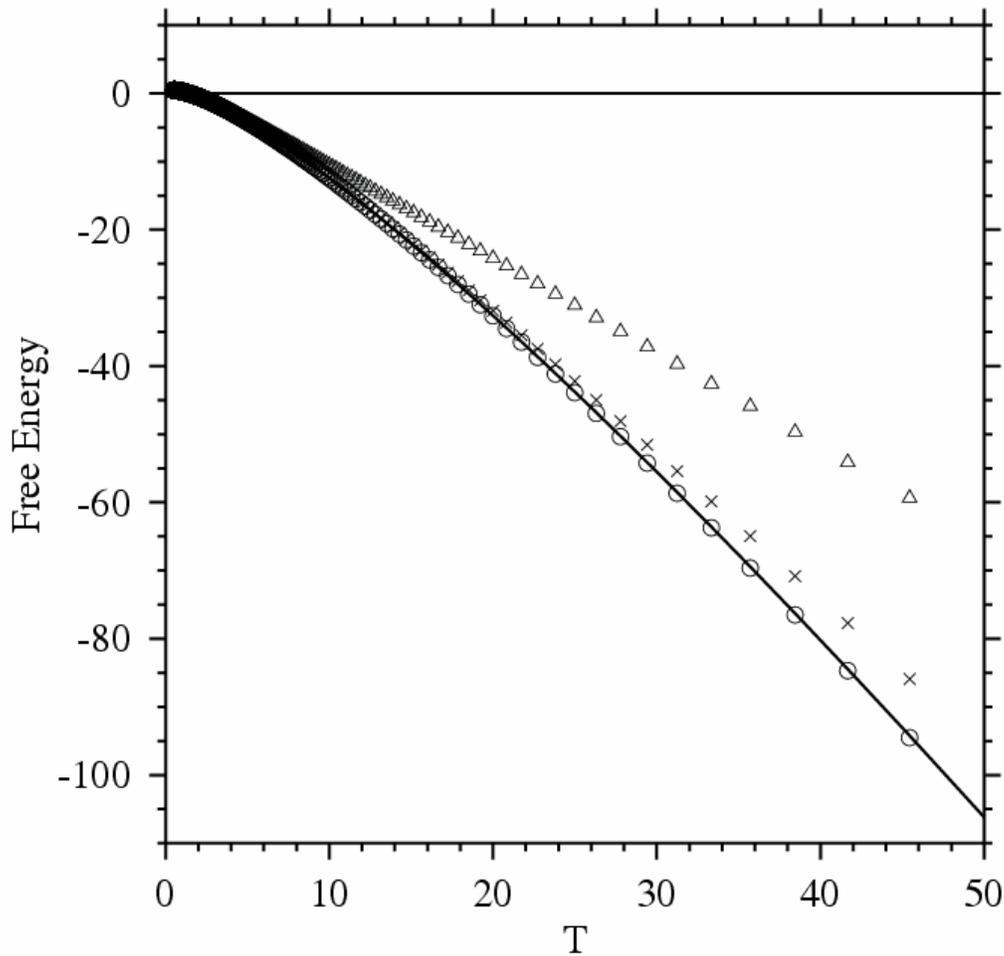

**Figure 3.** A comparison of numerical results for the free energy as a function of temperature with the exact solution (solid line) for a particle in an infinite square well. The temperature range considered is 0.5 to 45. The circles, crosses, and triangles correspond to 49, 9 and 4 initial wave functions respectively. As in Fig. 2, there is good convergence with respect to the number of low-energy sine functions in the initial wave function set. The low temperature limit of the free energy is the ground state energy $E_g = 1/2$. The numerical errors for circles are less than 0.4 %.

**4.2. One-dimensional harmonic oscillator**

We next apply the numerical method to a one-dimensional harmonic oscillator. This is a standard test case for a particle bound in a smooth potential well, in this case $V(x) = x^2/2$. The exact energy eigenvalues and eigenfunctions are well known and the exact Bloch density matrix given by Eq. (1) can also be evaluated analytically [17,25]

$$C(x, x', \beta) = \sqrt{\frac{1}{2\pi \sinh(\beta)}} \exp\left\{\frac{-1}{2\sinh(\beta)}\left[(x^2 + x'^2)\cosh(\beta) - 2xx'\right]\right\} \qquad (22)$$

For the numerical tests of our method in this case, the parameters $\Delta x = 0.2$ and $\Delta \tau = (\Delta x)^2/4 = 10^{-2}$ are used. In this and subsequent applications the size of the computational cell is chosen so that all wave functions have decayed essentially to zero, hence giving negligible truncation error. In this case a cell length of 20 was sufficient. As in the previous example, contour plots of the numerical and exact density matrix are in excellent agreement, so these plots are not repeated here. Numerical results for the free energy are compared with the exact results over the temperature range 0.125 to 50 in Fig. 4. The agreement is excellent. The numerical results (using 99 wave functions) agree to within less than 0.5% with the exact results over the entire range. Figure 4 also demonstrates how the number of initial functions required to obtain accurate results depends on the temperature range and that only a small number is needed at low $T$. The low temperature limit of the free energy is in agreement with the ground state energy $E_g = 1/2$ and the entropy approaches zero.

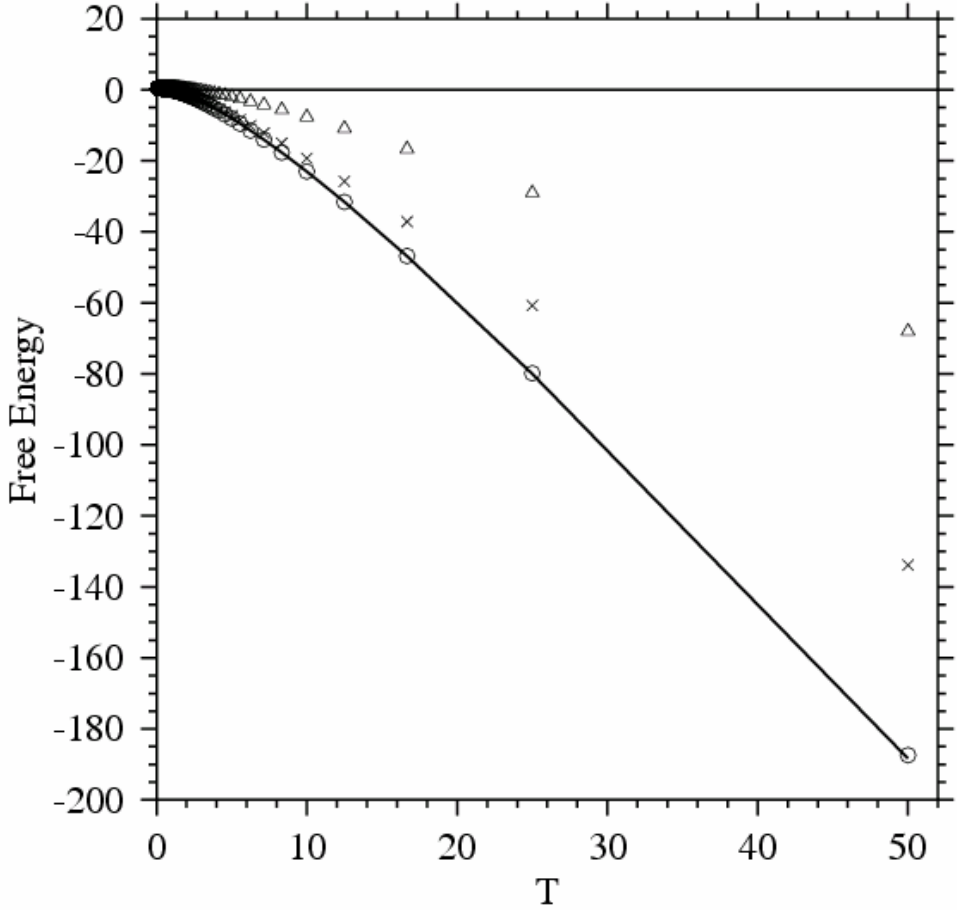

**Figure 4.** A comparison of numerical results (circles, crosses, and triangles) of the partition function with the exact solution (solid line) as a function of temperature for a particle in a harmonic oscillator for different number of initial sine wave functions. The temperature range considered is 0.125 to 50. The circles, crosses and triangles correspond to 99, 20 and 5 low-energy wave functions, respectively.

These two examples confirm the accuracy and convergence properties of our procedure in known test cases. We will next consider other applications where exact results for the density matrix and thermodynamic properties have not previously been known.

**4.3. One-dimensional linear potential**

The potential generated at the interface between a semiconductor and an insulator in quasi-two-dimensional nanostructure devices can confine the motion of electrons to the vicinity of the interface. Taking the insulating side of the interface to be impenetrable, an electron near the interface moves in a potential which can be approximated as linear as a function of the distance $x$ from the electron to the interface plane.

$$V(x) = \begin{cases} x & \text{if } x > 0 \\ \infty & \text{if } x < 0 \end{cases}. \tag{23}$$

Eigenfunctions and implicit results for energy eigenvalues are known analytically in terms of Airy functions[26].

We have computed the density matrix, free energy, internal energy and entropy for the motion of a particle in the linear potential Eq. (23) for temperatures in the range 0.2 to 4.0. The parameters were taken as $\Delta x = 0.1$ and $\Delta \tau = (\Delta x)^2 / 4 = 0.0025$ with a computational cell length of 30. The number of initial wavefunctions used is 299. Numerical results for the free energy are shown in Fig. 5. For comparison with the FDTD results we also computed the free energy using the first 60 eigenvalues obtained from zeros of Airy functions. It is shown in Fig. 5 that the numerical results from the two methods are in excellent agreement. The finite limit of the free energy extrapolated to $T = 0$ is found to be 1.85461 which is in good agreement with the value $E_g = 2.33810741 / \sqrt[3]{2} \approx 1.85576$ given in [26]. The computed entropy extrapolates smoothly to zero, as expected for the $T \to 0$ limit. To check the effect of the size of the grid mesh, we have recomputed using the parameters $\Delta x = 0.025$, $\Delta \tau = (\Delta x)^2 / 4$ with the same computational length of 30. The ground state energy as estimated from the limit of the free energy extrapolated to $T = 0$ is now found to be 1.85569. This clearly shows how the FDTD results can be systematically improved using smaller grid size.

**4.4. One-dimensional quartic oscillator**

Anharmonic oscillator potentials are ubiquitous in non-relativistic quantum systems and are also useful to illustrate some aspects of quantum field theory. General analytical solutions are not available for such problems and numerical methods are essential. The simplest example of such problems is the quartic oscillator with the potential energy given by $V(x) = x^4$.

To demonstrate the FDTD method for this one-dimensional quartic oscillator, we have computed the density matrix and thermodynamic properties using the parameters $\Delta x = 0.2$ and $\Delta \tau = (\Delta x)^2 / 4$ with a computational length of 20. The number of initial wave functions used here is 99. Numerical results for the free energy for temperatures in the range 0.2 to 4.0 are shown in Fig. 5. The exact density matrix for the quartic anharmonic oscillator is unknown although accurate energy eigenvalues and wave functions for the quartic oscillator are available. For comparison with the FDTD results for the same temperature range we also computed essentially exact thermodynamical properties using the first 50 eigenvalues given in [27]. The corresponding free energy is also shown in Fig. 5. The numerical results obtained by the two methods are clearly in good agreement. As in the previous examples, the free energy approaches the ground state energy as the temperature approaches zero and the entropy approaches zero. Our numerical ground state energy is found to be $E_g = 0.66423$ in close agreement with the value $1.060362 / 2^{2/3} \approx 0.667986$ given in [27]. Similar as in previous section, by reducing the grid spacing to $\Delta x = 0.05$, we obtain an improved ground state energy $E_g = 0.66775$.

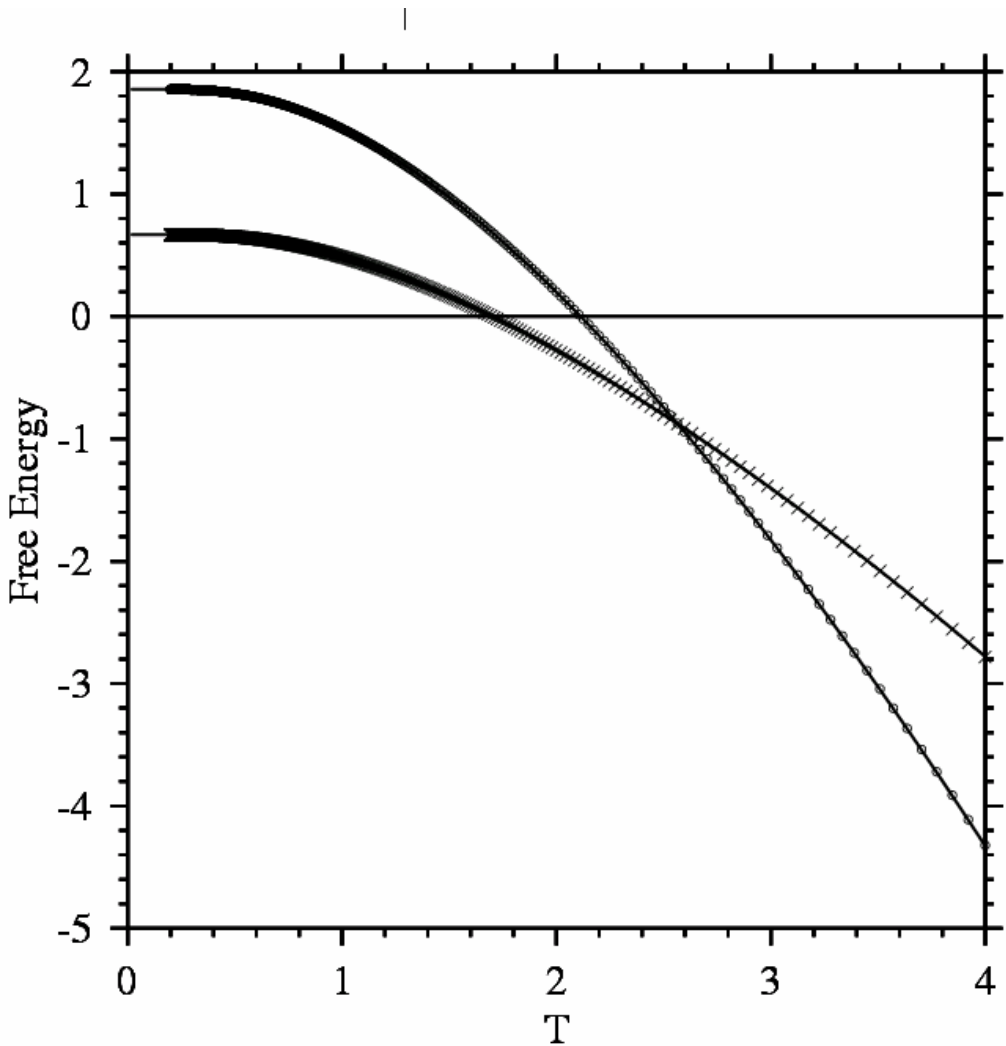

**Figure 5.** Free energy F as a function of temperature for a particle in the linear potential given by Eq. (23) (circles) and in a quartic oscillator potential $V(x) = x^4$ (crosses) computed using the FDTD method. The temperature range considered is 0.2 to 4.0. The solid line curves are for the accurate results computed using 60 eigenvalues (the zeros of Airy functions) for the linear potential and using 50 eigenvalues given in [27] for the quartic potential.

**4.6 One-dimensional quartic double well potential**

The previous power law examples illustrate the FDTD method in cases where the one-body potentials all have a single local mimimum (taken to be at $x = 0$) within the computational domain. There are many interesting situations where the one-body potential has two (or more) local minima which are degenerate or very nearly so. This type of model potential can describe important physical aspects of structural phase transitions, tunneling of protons in hydrogen-bonded systems, and other systems with inversion symmetry such as $NH_3$ molecules. The degeneracy may occur for symmetry reasons (as in the case of an $NH_3$ molecule) or by design (an example is given in subsection 4.7).

The degeneracy in such cases introduces an important new physical feature which must be taken into account. This can be clearly illustrated by a single particle in a one-dimensional quartic double well potential given by

$$V(x) = (x^2 - a^2)^2 \ . \tag{24}$$

The two minima of this double well potential are located at $\pm a$ and the barrier centered at $x = 0$ is of height $a^4$. For very large $a$ the energy spectrum associated with Eq. (24) corresponds approximately to two uncoupled degenerate harmonic oscillators. It is well known that this degeneracy is lifted for any finite value of $a$. In particular, the ground state is unique and the splitting $\Delta = \Delta(a) = E_2 - E_1$ between the ground state and the first excited state is always finite, albeit possibly small for large $a$. It is important to see how this energy splitting scale is reflected in thermodynamical properties, especially at low $T$.

We have computed the density matrix and thermodynamical properties for $a = 1.0, 1.2, 1.4$ and $2.0$, and for $T$ in the range $0.02$ to $4.0$. In this application, we used the parameters $\Delta x = 0.1$, $\Delta \tau = (\Delta x)^2 / 4$ with a computational cell length of 10.

For all values of the parameter $a$, the free energies and internal energies were found to approach the ground state energies smoothly as $T \to 0$. This is as expected so their graphs will not be shown. However the potential barrier leads to striking variation in the slope of the free energy at low $T$. The entropy $S(a, T) = -\partial F / \partial T$ is plotted in Fig. 6 for $T$ in the range $0.02$ to $4.0$. The expected limiting value of zero for $S(a, T) \to 0$ is clear for $a = 1.0, 1.2$ and $1.4$, although the approach to zero is seen to be delayed substantially as $a$ increases. In fact, for $a = 2$ the entropy has only reached a constant plateau value of $\ln(2)$ in the computed range of $T$ and shows no indication of decreasing to zero. However, it would be incorrect to interpret this plateau value as a "residual entropy" since the lowest temperature plotted in Fig. 6 is only $T = 0.02$.

It is clear from Fig. 6 that determination of the low $T$ limit of the entropy by extrapolation can be delicate in cases where the confining potential has degeneracies. If the entropy does not appear to extrapolate to zero in any particular computation, it is always necessary to determine whether the low $T$ limit has in fact been reached. In other words, is there any energy scale in the problem which is below the computed range? We emphasize that all of the computed data shown in Fig. 6 are correct. In this double well model the low $T$ limit where $S(a, T) \to 0$ will be observed clearly requires $T \ll \Delta(a) = E_2(a) - E_1(a)$. In the case of high potential barrier, the splitting is small relative to the ground state energy so we will first use a semi-classical method to obtain an analytical expression for the energy splitting. The dependence of $\Delta_{SC}(a)$ on $a$ is [28]

$$\Delta_{SC}(a) \approx 16 \, 2^{1/4} \pi^{-1/2} \, a^{5/2} \, \exp[-4\sqrt{2}a^3 / 3] \tag{25}$$

Note that the dominant dependence by far at large $a$ is due to the $a^3$ factor in the exponential. This semi-classical estimate gives $\Delta_{SC}(a=2) \approx 1.7064x10^{-5}$ which is far below the minimum temperature $T = 0.02$ for the data in Fig. (6). Clearly the "low $T$" limit has not been approached for $a = 2$ and the plateau value of the entropy apparent in Fig. 6 simply corresponds to the apparent double degeneracy of the ground state when the level splitting is not resolved.

It is useful to check the accuracy of the semiclassical estimate and to verify that the interpretation is consistent. A direct computation of energy eigenvalues using the FDTD method [12] gives $\Delta(a=2) \approx 1.5901x10^{-5}$ so the semiclassical result is an acceptable estimate for $a = 2$. Note that the semi-classical estimate is valid only for large value of $a$. For $a = 1$, the semiclassical estimate using Eq. (25) is $\Delta_{SC}(a=1) \approx 1.6288$ while the computed FDTD eigenvalues [12] give $\Delta(a=1) \approx 0.7918$. In this case the semiclassical estimate is only accurate within a factor of two but has still correctly indicated a splitting temperature scale of order unity for $a = 1$. This is consistent with Fig. (6).

We see from this double well example that the FDTD method for computing the density matrix and thermodynamic properties accurately describes cases where the one-body potential has degenerate local minima. Of course, as clearly shown by the four cases in Fig. 6, in order to exhibit the effect of this degeneracy the computed range of $T$ considered must contain the temperature scale by which the classical degeneracy is split. In particular, the low temperature limit requires $T$ small with respect to any splitting scales. Ultimately the third law of thermodynamics is satisfied and $S(a,T)$ will always approach zero when $T/\Delta(a) \to 0$ *provided* the system is in thermal equilibrium.

The determination of whether or not a system under observation is in thermal equilibrium is a separate issue but a brief comment is in order. In general, a system prepared in an initial state will closely approach equilibrium when the observational time is longer than the relevant relaxation times of the system. In this double well example, the time $\tau \sim 1/\Delta$ for oscillations (or tunneling) between the two wells is a relevant internal time scale for very low $T$ processes. Even for this simple example we see that $\tau$ can vary by orders of magnitude when the barrier height increases.

It should be noted that some physical materials have very long relaxation times, especially at low temperature, so that thermal equilibrium is actually not established on realistic experimentally accessible time scales. One celebrated example is the residual entropy experimentally observed for common ice (phase Ih). The entropy associated with proton disorder on hydrogen bonds can be approximately described by a double well model for each bond. Taking into account the constraints of the three-dimensional network of $H_2O$ molecules, Pauling estimated the residual entropy of ice (in units of $k_B$) to be $S = \ln(4/3)$ per water molecule[29]. This estimate is already in quite good agreement with the experimental value and corrections have also been calculated. References and discussion are given in [30]. Similar residual entropy is observed in macroscopic glassy systems.

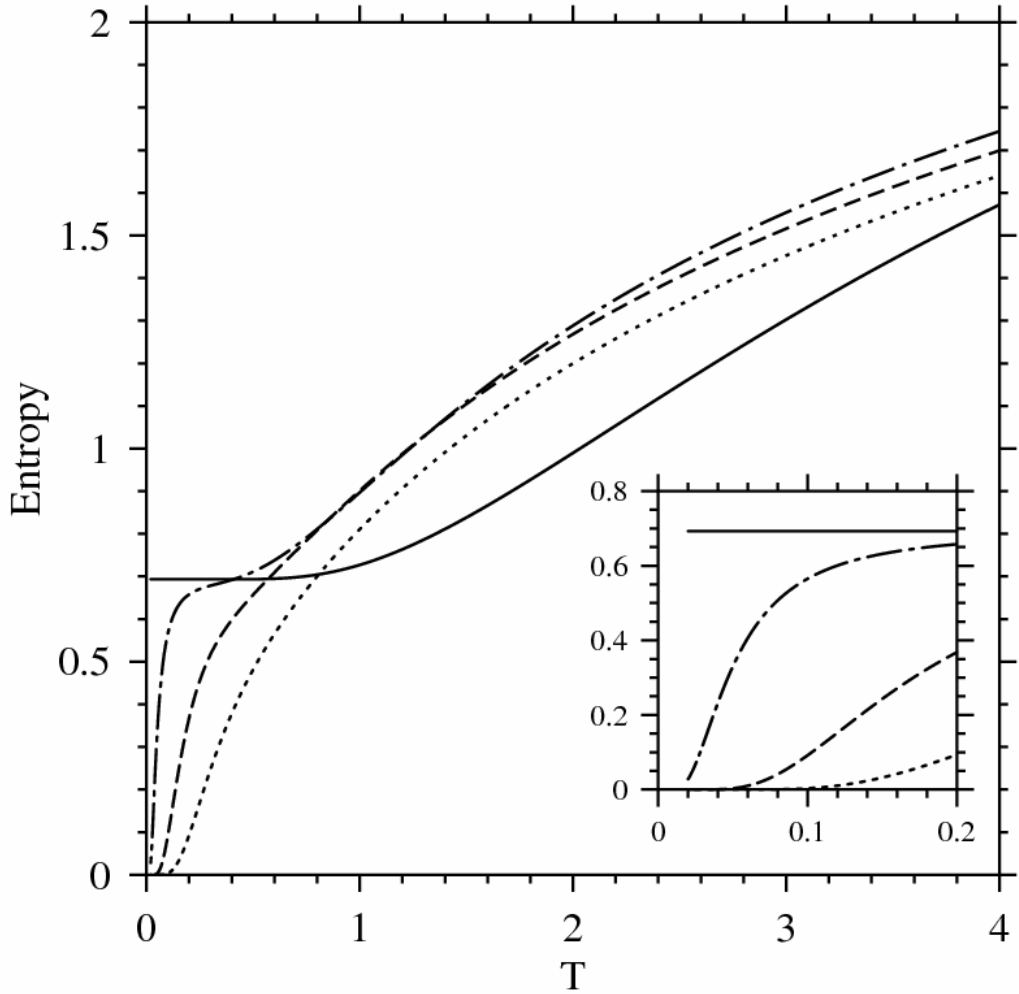

**Figure 6.** Entropy as a function of temperature for a particle in double well potential $V(x) = (x^2 - a^2)^2$ for three different barrier parameters: $a = 1$ (dotted line), $a = 1.2$ (dashed line), $a = 1.4$ (dash-dotted line), $a = 2$ (solid line). The wide temperature range shown is 0.02 to 4.0. The inset shows the entropy in the lower temperature range 0.02 to 0.2. The correct low temperature limit of zero is evident except for the $a = 2$ case. However, the data shown are all correct. In the $a = 2$ case, the results are also correct for the indicated range. Computations at much lower temperature, $T \sim \Delta(a = 2) \sim 10^{-5}$ on the scale of the degeneracy splitting, would reveal the approach to the ground state limit (see text).

**4.7. Two-dimensional double quantum dots**

Quasi-two-dimensional electronic systems with only a few electrons play an increasingly important role in a wide range of modern nanostructure devices. In this section we apply our FDTD procedure to study thermodynamic properties of a two-dimensional double quantum dot containing one electron. We are particularly interested in the response of the entropy and the electron density to an applied electric field. To model a double quantum dot with an applied electric field, we use a potential given by Burkard et al. [31]

$$V(x, y) = \frac{1}{2}\left[\frac{(x^2 - a^2)^2}{4a^2} + y^2\right] + eEx \tag{26}$$

The first term represents two fused quantum dots separated by a potential barrier centered on the $x = 0$ plane. There are two local minima located at $x = \pm a, y = 0$ in the absence of the field. The second term describes the electric field applied in the positive x direction, $-e$ is the electron charge and $E$ is the magnitude of the field. The factor $1/(4a^2)$ in Eq. (26) is included so that the potential is locally an isotropic harmonic oscillator at $x = \pm a, y = 0$ when $E = 0$.

For the computations, the parameters $a = 4$, $\Delta x = \Delta y = 0.2$ and $\Delta \tau = (\Delta x)^2 / 8$ are used. The computational cell was taken to have side length of 20 in the x direction and 10 in the y direction. In this case we use $49 \times 99 = 4851$ initial wave functions. We computed the density matrix $\rho(x, y; x', y'; \beta)$ and thermodynamic properties for a range of values of $T$ and $E$. Numerical results for the entropy are shown in Fig. 7 for $T$ in the range 0.005 to 0.5 and for $eE = 0, 0.02$ and $0.1$. The results for zero electric field are similar to the previous $a = 2$ one-dimensional case. The entropy exhibits an apparent "residual" plateau value of $\ln(2)$ at low temperature down to $T = 0.005$. This indicates that the energy splitting $\Delta$ between the ground state and the first excited state in this double quantum dot system is much less than $0.005$ so the two quantum dots appear to be degenerate (like the previous example when $a$ is large). The applied electric field breaks inversion symmetry along the x-axis and removes this degeneracy. The net effect can be thought of as introducing an energy gap of order $2eEa$ which is much larger than the intrinsic $\Delta$ for the finite values of electric fields used here. Then the apparent residual entropy is removed and entropies in a finite field clearly extrapolate to zero on the scale of Fig. 7. There is no doubt that $S(T \rightarrow 0) = 0$.

It is interesting to see the effect of the applied electric field on the particle density distribution. We define the integrated density $n(x)$ by integrating the two-dimensional density $n(x, y)$ with respect to the y variable. The integrated particle density is plotted as a function of $x$ in Fig. 8 for $T = 1$. By symmetry, $n(x)$ is even in $x$ in the absence of an applied field, irrespective of $T$ versus $\Delta$, when the system is in thermal equilibrium. Fig. 8 shows that this spatial symmetry is broken by the applied electric field. The particle density becomes strongly asymmetric and depends on both the direction and strength of the electric field.

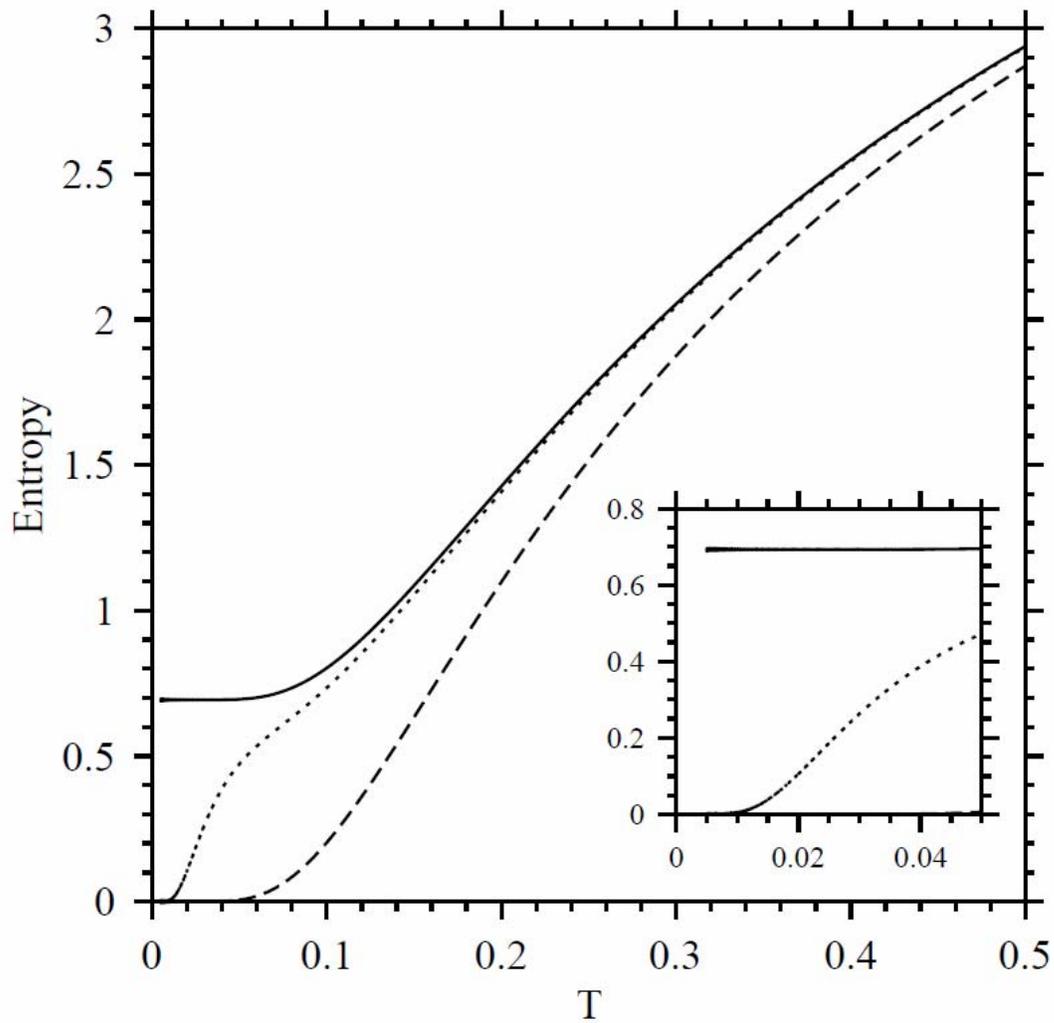

**Figure 7.** Entropies as a function of temperature (in energy units) for a particle in a double quantum dot with potential given by Eq. (26) for $a = 4$ and for three different values of the electric field parameter: $eE = 0$ (solid line), $eE = 0.02$ (dotted line), $eE = 0.1$ (dashed line). The wide temperature range shown is 0.005 to 0.5. The inset shows the data for $eE = 0$ and $eE = 0.02$ plotted in the lower range 0.005 to 0.05. The intrinsic energy splitting $\Delta$ between the ground state and the first excited state is much less than $0.005$. The inset shows that the low temperature limit has been approached for $eE = 0.02$ but not for $eE = 0$.

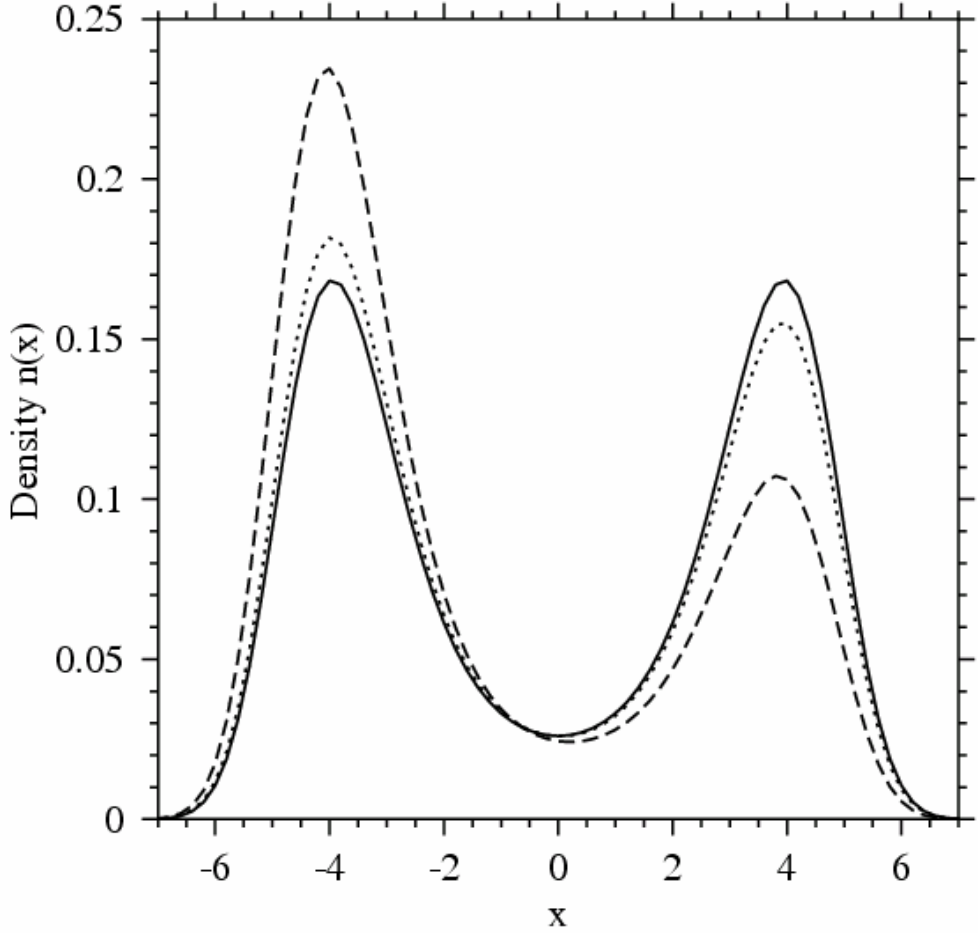

**Figure 8.** Integrated particle densities distribution for a particle in a double quantum dot with potential given in Eq. 26 for $a = 4$ and temperature $T = 1$ and for three different values of the electric field parameter: $eE = 0$ (solid line), $eE = 0.02$ (dotted line), $eE = 0.1$ (dashed line). The integrated densities are obtained by integrating the two-dimensional densities $n(x, y)$ with respect to the y variable.

**4.8. Three-dimensional potentials**

Finally we apply our FDTD procedure to three-dimensional problems. We considered harmonic and quartic oscillators with similar conclusions. We make no use of symmetries in the computations and other three-dimensional problems with a general potential can be solved without difficulty using this numerical method. Results will be given here only for the quartic oscillator with $V(x, y, z) = (x^4 + y^4 + z^4)$. The numerical results for free energies, internal energies and entropies are given in Fig. 9 for temperatures in the range 0.1 to 1.7. These results were computed using the parameters $\Delta x = \Delta y = \Delta z = 0.2$ and $\Delta \tau = (\Delta x)^2 / 20$. The computational volume was taken to have side length of 20. Using these parameters there are about $100 x 100 x 100 = 10^6$ grid points needed to be stored in memory. If double precision is used, the amount of memory needed is about 8 MBytes. Since we have to store previous values of variables in the iteration, we need to store twice of this amount of memory. In contrast, if the DPI method is used we would need about 8000 GBytes of memory.

Numerical results using 8000 initial functions are given in Fig. 9 for the free energy, the internal energy, and the entropy. The CPU time needed to compute the partition function at temperature $T = 1$ (250 iteration steps) using one initial wave function on an Intel Pentium 4 1.8 GHz computer is 12.55 seconds. Therefore for 8000 initial wave functions, we need about 27 hours of CPU time. For comparison, the CPU time needed for the corresponding two-dimensional systems at $T = 1$ (250 iteration steps) using 400 initial wave functions is about 25 seconds. Similarly, the CPU time needed for the corresponding one-dimensional systems at $T = 1$ (250 iteration steps) using 20 initial wave functions is about 0.01 second. In all cases the CPU times are proportional to the number of initial wave functions so it is obvious that this FDTD method is ideal for computing in parallel.

A direct numerical check on the accuracy of these results is available because variables in this quartic potential are separable. Then the resulting thermodynamical properties of the three-dimensional system can also be obtained using products of partition functions of the corresponding one-dimensional problem treated in section 4.4. It is seen in Fig. 9 that the numerical results obtained using the full three-dimensional computation are in excellent agreement with the products of one-dimensional partition functions. We verify that the ground state energy of the three-dimensional quartic potential (1.9940) is equal to three times the corresponding one-dimensional quartic potential ground state energy.

This three-dimensional example emphasizes two very useful features of the FDTD method. First, it is sufficient to use only a relatively small number of initial wave functions to get accurate results for finite temperature systems. In the present case the grid allows for a maximum of about $10^6$ initial functions. In our computations only $8000$ initial functions were sufficient to produce accurate results within 1%. This corresponds to $(8000)^{1/3} = 20$ initial functions per dimension which is comparable to previous examples. The numerical results can always be easily improved by using more initial functions. Second, the separability of the potential in this example played no role in the computations. This separable $V(x, y, z)$ was used only to give a convenient numerical check. The computations can be carried out with equal ease for general potentials without any particular symmetries.

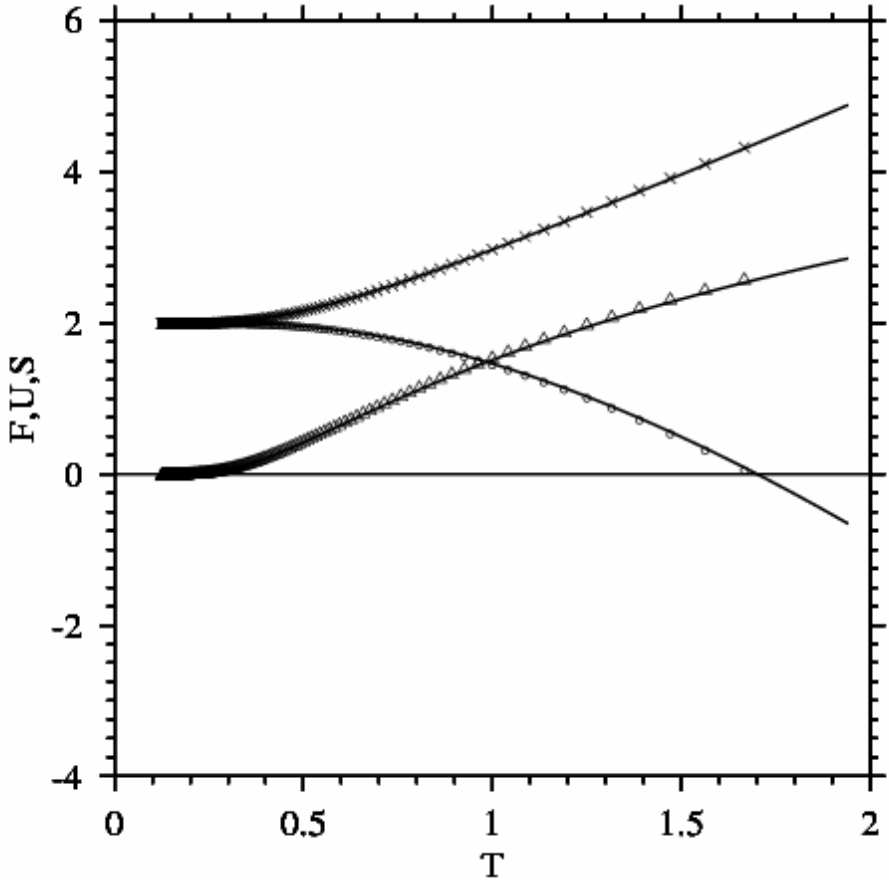

**Figure 9.** Free energy (circles), internal energy (crosses) and entropy (triangles) as a function of temperature for a particle in three-dimensional quartic oscillator $V(x, y, z) = (x^4 + y^4 + z^4)$. Solid line curves are computed using products of the partition functions of the one-dimensional potential $V(x) = x^4$ given in section 4.4. The temperature range shown is 0.1 to 1.7.

## 5. Conclusions

A general method for computing the thermal density matrix and thermodynamic properties of a single-particle system has been presented. The accuracy of the method was verified by examples for which exact analytical results exist for comparison. New numerical results have been given for linear and quartic single well potentials, and for a quartic double well potential in one dimension. Results have been given for a double quantum dot in two dimensions, including an applied electric field, and for a quartic anharmonic oscillator in three dimensions. These examples illustrate the efficiency and accuracy of the method for a wide variety of one-body potentials.

We summarize several important useful features of our procedure. (1) No information on the exact eigenvalues or eigenfunctions is required. Instead a set of initial functions evolving in imaginary time is used to construct the density matrix. Integration of the Schrodinger equation in imaginary time is carried out by the finite difference time domain method. The number of functions to be used in this initial set depends on the required accuracy which is specified by the user. (2) The procedure is well suited for parallel computation since the iterations for every initial wave function can be performed independently. (3) The FDTD method is easily implemented and uniformly applicable for general confining potentials and spatial dimension. We expect it to be useful in practical finite temperature applications to quasi-two-dimensional nanostructures and other low symmetry quantum systems.

The FDTD method for systems in a variety of inhomogeneous confining potentials was applied here for the single-particle case $N = 1$. The theoretical framework itself applies much more generally. We have verified that the numerical procedure can be extended to systems of two or more particles, *including* the inter-particle interactions. Note that since our FDTD method deals directly with wave functions, the anti-symmetric property of wave functions for identical fermions can be incorporated exactly. There is no sign problem in the FDTD method. Finally, we considered here only cases where the wave functions can be taken to be real. Following [20] for the single-particle case, we have also extended the FDTD procedure for interacting systems to include applied magnetic fields. Further details and applications of these results will be given in a future paper.

## Acknowledgements

Part of this work was done while one of the authors (DJWG) was attending a workshop on many-body theory at Centro di Ricerca Matematica "Ennio De Giorgi", Pisa, Italy. The hospitality of the Centro De Giorgi as well as discussions with N.H. March are acknowledged.